\documentclass[preprint,sort&compress,11pt,titlepage]{elsarticle}

 \usepackage{graphicx}
\usepackage{amssymb}
\usepackage{amsthm}
\usepackage{amsmath}

\begin{document}

\newtheorem{remark}{\textbf{Remark}}
\newtheorem{theorem}{\textbf{Theorem}}
\newtheorem{definition}{\textbf{Definition}}
\newtheorem{proposition}{\textbf{Proposition}}
\newtheorem{corollary}{\textbf{Corollary}}

\begin{frontmatter}



\title{Density-based Monte Carlo filter and its applications in estimation of unobservable variables and pharmacokinetic parameters}


\author[a,c]{Guanghui Huang}
\ead{hgh@cqu.edu.cn}

%\author[c]{Li Cao}

\author[b]{Jianping Wan}
\ead{hust\_jp\_w@yahoo.com.cn}

\author[a]{Hui Chen \corref{cor1}}
\ead{chenhui\_tj@126.com}

\cortext[cor1]{Corresponding author.}

\address[a]{
Department of Pharmacology, Tongji Medical College, Huazhong University of Science and Technology, Wuhan 430030, China}

\address[b]{College of Mathematics and Statistics, Huazhong University of Science and Technology, Wuhan 430074, China}

\address[c]{College of Mathematics and Statistics, Chongqing University,
  Chongqing 401331, China}

\begin{abstract}

Nonlinear stochastic differential equation models with unobservable variables are now widely used in the analysis of PK/PD data.
The unobservable variables are often estimated with extended Kalman filter (EKF), and the unknown pharmacokinetic parameters are usually estimated by maximum likelihood estimator. However, EKF is inadequate for nonlinear PK/PD models, and MLE is known to be biased downwards. A density-based Monte Carlo filter (DMF) is proposed to estimate the unobservable variables, and a simulation-based procedure is proposed to estimate the unknown parameters in this paper, where a genetic algorithm is designed to search the optimal values of pharmacokinetic parameters.  The performances of EKF and DMF are compared through simulations,  and it is found that the results based on DMF are more accurate than those given by EKF with respect to mean absolute error.

\end{abstract}

\begin{keyword}
PK/PD modeling \sep Stochastic differential equation \sep Extended Kalman filter \sep Density-based Monte Carlo filter \sep Genetic algorithm

%JEL: G13
\end{keyword}

\end{frontmatter}

\titlepage

%%\linenumbers

\section{Introduction}
Stochastic differential equations (SDEs) are  powerful tools in pharmacokinetic and pharmacodynamic (PK/PD) modeling,
which can be used as a diagnostic tool to facilitate systematic model development \citep{kristensen2005,tornoe2005,overgaard2007} or as a realistic method to describe the variations in system \citep{andersen2005,ditlevsen2005,overgaard2005,moller2010,picchini2010}.
\citep{overgaard2007} documents that SDEs provide a more realistic description of the variability that improves individual simulation and predictive properties, accelerates model speed by simplifying inter-occasion variability, and finally changes the model into one that could not be falsified by the autocorrelation function.

Pharmacokinetic parameter estimation is one of the important steps in PK/PD  data analysis. Maximum likelihood estimation (MLE) based on the extended Kalman filter (EKF) is usually applied to estimate  the parameters in SDE models, such as  \citep{kristensen2005,tornoe2005,overgaard2007,overgaard2005,moller2010,picchini2010} and the references therein. On the other hand, Kalman filter is designed to estimate the state variable involved in a linear model, and EKF is the linearized version of Kalman filter for the models with nonlinear characteristics. \citep{overgaard2007} documents that the failure to produce Gaussian residuals with EKF may indicate which is inadequate for nonlinear modeling in PK/PD data analysis, possibly motivating the pursuit of higher order filters or other estimation methods.

\citep{tanizaki1996} argues  that even if the higher-order nonlinear filters deduced from Kalman filter give us less biased filtering estimates than the EKF, the filtering estimates obtained from the higher-order nonlinear filters are still biased because the nonlinear functions are approximated ignoring the other higher-order terms. And the other filters, such as the density-based Monte Carlo  filters (DMF), might be less biased than the EKF, as the unobservable variable can be generated from the nonlinear functions directly without approximations.  The purpose of this paper is to compare the performances of EKF and DMF under a nonlinear model, and to develop more efficient algorithms for PK/PD parameter estimation.

As MLE is often inefficient and biased for finite sample \citep{vaart1996,meza2007}, a simulation-based procedure is proposed to estimate the unknown parameters in this paper, where a genetic algorithm is designed  to search the optimal values of parameters.  A one-compartment pharmacokinetic model with nonlinear absorption and first order elimination is used to compare the performances of EKF and DMF through simulated investigations.
It is found that the results based on DMF is more accurate than those based on EKF with respect to mean absolute error.

The remainder of the paper is constructed as follows: Section 2 introduces the model to be investigated, Section 3 gives the EKF algorithm for this model, and Section 4 demonstrates the proposed DMF algorithm. The estimates of the unobservable variables by EKF and DMF are compared in Section 5. The criterion of estimation for the unknown parameters and the genetic algorithm of optimization are given in Section 6. The conclusions and discussions are given in Section 7.

\section{Stochastic nonlinear model}
A one-compartment model with nonlinear absorption and first order elimination is considered in this paper, which is used to describe the PK of a drug following an oral dose by \citep{kristensen2005}, where pharmacokinetic parameters are estimated with MLE based on EKF.

Let $Q(t)$ (mg) be the amount of drug in the GI tract at time $t$, $t\in (0,T]$ (min), which is an unobservable  variable.
$C(t)$ (mg{/}l) is the concentration of drug in plasma, which is an observable variable. A system of stochastic differential equations are used to describe the processes of absorption and elimination of the drug, i.e.
\begin{eqnarray}
\mathrm{d}Q(t)
 & = &
  -\frac{V_{max}~ Q(t)}{K_{m} + Q(t)} \mathrm{d}t + \sigma_q \mathrm{d}B(t),\label{state}\\
\mathrm{d}C(t)
 & = &
  \left[ \frac{V_{max}~ Q(t)}{\left(K_{m}+Q(t)\right)~ V}
   - \frac{C_{L} C(t)}{V} \right] \mathrm{d}t
    + \sigma_c \mathrm{d}W(t), \label{observation}
\end{eqnarray}
where $V_{max}$ (mg/min) is the maximum reaction rate, $K_{m}$ (mg) is the Michaelis constant,
$\sigma_q$ and $\sigma_c$ are the diffusion parameters of $Q(t)$ and $C(t)$ respectively, $C_{L}$ (l/min) is the rate of elimination, and $V$ (l) is the apparent volume of distribution. The two stochastic processes $B(t)$ and $W(t)$, $t\in [0,T]$, are two independent standard Wiener processes starting from zero.
Let $\theta=\left( V_{max}, K_{m}, V, C_{L}, \sigma_q^2, \sigma_c^2 \right)$, which is a vector of six elements, and  $\theta \in \Theta \subset \mathbb{R}^6$, where $\Theta$ is the parameter space. The purpose of this paper is to estimate $\theta$ from the limited observations of $C(t)$, $\left\{c_{t_1}, c_{t_2}, \cdots, c_{t_n} \right\}$, where $\left\{t_1,t_2,\cdots,t_n \right\}$ are the time points of observations, and  $n$ is the number of observations.

SDEs (\ref{state}) and (\ref{observation}) are nonlinear equations, it is difficult to find explicit solution for such SDEs. In order to simulate $Q(t)$ and $C(t)$ at discrete time points,  (\ref{state}) and (\ref{observation}) are approximated with discrete differences in It{\^o} type, i.e.
\begin{eqnarray}
Q_{k} &\cong &
  Q_{k-1} - \frac{V_{max}~Q_{k-1}}{K_{m} + Q_{k-1}}\left( t_{k}-t_{k-1}\right)
    + \sigma_{q} \left( B_{k} - B_{k-1} \right), \label{discretestate}\\
C_{k} & \cong &
   C_{k-1} + \left[ \frac{V_{max}~Q_{k-1}}{\left(K_{m}+Q_{k-1}\right)~V}
     -\frac{C_{L}~C_{k-1}}{V}
     \right]
      \left( t_{k} - t_{k-1} \right)
      +\sigma_c \left(W_k - W_{k-1}\right),  \label{discreteobserved}
\end{eqnarray}
where $Q_{k}=Q_{t_k}$, $C_{k} = C_{t_k}$, $B_k = B_{t_k}$ and $W_k = W_{t_k}$, $k=1,2,\cdots, n$.  $B(t)$ and $W(t)$ are two independent standard Wiener processes, such that the increments $B_k - B_{k-1}$ and $W_k - W_{k-1}$ are independent and identically distributed, where $B_k - B_{k-1} \sim N(0, \sqrt{t_k - t_{t-1}} )$.

\section{Extended Kalman filter}

Let $c_k=c_{t_k}$, which is the observation at time $t_k$.
And the information set at time $t_s$ is $Y_s = \left\{ c_1,c_2,\cdots,c_s \right\}$, where $s \in \left\{ 1,2,\cdots,n \right\}$. Let
\begin{equation}\label{filters}
Q_{k|s}=\mathrm{E} \left[ Q_k | Y_s \right],
\end{equation}
which is the conditional expectation of $Q_{k}$ given the information set at time $t_s$. When $s<k$, (\ref{filters}) is called the prediction of $Q_k$; when $s=k$, (\ref{filters}) is called  the filtering of $Q_k$; and (\ref{filters}) is called the smoothing of $Q_k$ when $s>k$.

Kalman filter is particularly powerful and useful for the linear models which include  unobservable components. Applying the linearized nonlinear functions to the Kalman filter, the resulted algorithm is called the extended Kalman filer (EKF).
Approximate the two nonlinear functions in (\ref{discretestate}) and (\ref{discreteobserved}) with first order Taylor series expansion, we have the following EKF algorithm ( see \ref{appekf}):
\begin{eqnarray}
Q_{k|k-1} &=&
   Q_{k-1|k-1} - \frac{V_{max}~Q_{k-1|k-1}}{K_{m}+Q_{k-1|k-1}}\left(t_{k}-t_{k-1}\right),\\
\Sigma_{k|k-1} &=&
   T_{k|k-1}^2 \Sigma_{k-1|k-1} + \sigma_q^2 \left( t_k - t_{k-1} \right), \\
C_{k|k-1} & =&
  C_{k-1} + \left[\frac{V_{max}~Q_{k-1 | k-1}}{\left( K_{m}+Q_{k-1 | k-1} \right)~V}
      - \frac{C_L ~ C_{k-1}}{V}\right] \left( t_k - t_{k-1} \right),\\
F_{k|k-1} &=&
  Z_{k|k-1}^2 \Sigma_{k|k-1} + \sigma_c^2 \left( t_k - t_{k-1} \right),\\
M_{k|k-1} &=&
  Z_{k|k-1} \Sigma_{k|k-1}, \\
K_{k} &=&
  M_{k|k-1} F^{-1}_{k|k-1},\\
\Sigma_{k|k} &=&
  \Sigma_{k|k-1}-K_{k}^2 F_{k|k-1},\\
Q_{k|k} &=&
  Q_{k|k-1} + K_{k} \left( C_{k} - C_{k|k-1} \right),
\end{eqnarray}
where
\begin{eqnarray}
Z_{k|k-1} & = &
  \frac{V_{max}~K_{m}}{\left( K_m + Q_{k-1 | k-1} \right)^2 ~ V} \left(t_{k}-t_{k-1}\right),\\
T_{k|k-1} &=&
 1- \frac{V_{max}~K_m}{\left( K_m + Q_{k-1 | k-1} \right)^2} \left(t_{k}-t_{k-1}\right).
\end{eqnarray}
Set $Q_{0|0}=Q_0$, $\Sigma_{0|0}=0$, $C_0 = 0$, the unobservable variable $Q_k$ can be estimated by EKF algorithm in a recursive manner.

\section{Density-based Monte Carlo filter}

Density-based Monte Carlo filter is an alternative solution to  nonlinear filtering problems, and the resulted algorithm is easy and convenient to compute the filtering estimate $Q_{k|k}$ \citep{tanizaki1996}.
The filtering estimation based on Monte Carlo technique is given by
\begin{equation}
Q_{k|k} = \sum_{j=1}^{N} Q_{j,k}~ \omega_{j,k} ,
\end{equation}
where $Q_{j,k}$ is the simulated value of the unobservable variable at time $t_k$ in the $j$th path, which is generated from equation (\ref{discretestate}) directly,  and $N$ is the number of simulated paths. $\omega_{j,k}$ is the weight of the $j$th path at time $t_k$, which satisfies $\omega_{j,0}=1/N$ and
\begin{equation}
\sum_{j=1}^{N} \omega_{j,k} = 1.
\end{equation}
$\omega_{j,k}$ is calculated with a recursive formula
\begin{equation}
\omega_{j,k}=
  \frac{p\left( c_k | c_{k-1}, Q_{j,k-1} \right) \cdot \omega_{j,k-1}}
       {\sum_{j=1}^{N}p\left( c_k | c_{k-1}, Q_{j,k-1} \right) \cdot \omega_{j,k-1}},
\end{equation}
where $p\left( c_k | c_{k-1}, Q_{j,k-1} \right)$ is the conditional density function of $C_k$ given by (\ref{discreteobserved}), i.e.
\begin{equation}
p\left( c_k | c_{k-1}, Q_{j,k-1} \right)
 =
  \frac{1}{\sqrt{2\pi}\sigma_k}
   \exp\left\{- \frac{\left( c_k - m_k \right)^2}{2 \sigma_k^2}  \right\},
\end{equation}
where
\begin{eqnarray}
\sigma_k^2
  & = &
    \sigma_c^2 \left(t_k - t_{k-1} \right),\\
m_k
  &=&
  c_{k-1} + \left[
             \frac
                 {V_{max}~Q_{j,k-1}}
                  {\left(K_{m}+Q_{j,k-1}\right)~V}
                  -
             \frac{C_{L}~c_{k-1}}
                  {V}
                  \right]
            \left( t_k - t_{k-1} \right).
\end{eqnarray}
Details of DMF can be found in \ref{appdmf}.

\section{Estimates of drug in GI tract}

$Q_{k}$ is the amount of drug in the GI tract at time $t_k$,  which is an unobservable variable when $t_k \in (0,T]$. The concentrations of drug in plasma can be observed at different time points, and $Q_k$ can be estimated from those observations by EKF and DMF respectively. In order to compare the performances of EKF and DMF, a simulated investigation is designed in this paper.

\begin{figure}[h]
  \caption{The observed concentrations in plasma
   and the estimated amounts of drug in GI tract.}\label{estimationqq}
\begin{center}
  \includegraphics[width=0.8\textwidth]{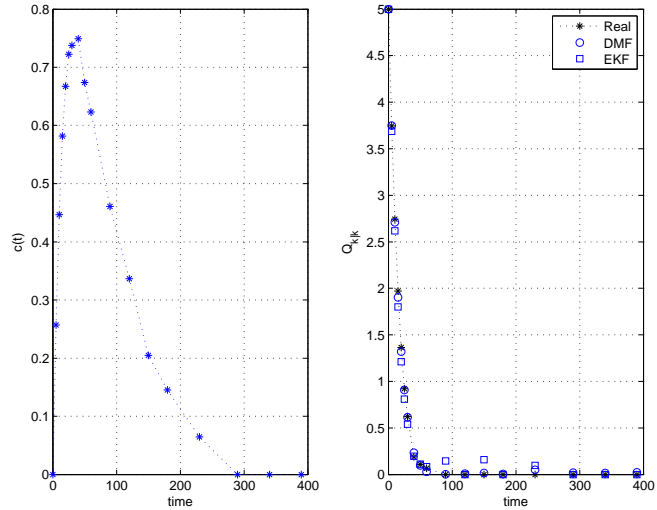}\\
\end{center}

\end{figure}

Set $Q_0 = 5$ mg, $c_0 = 0$ mg/l, $C_L = 0.05$ l/min, $V=5$ l, $V_{max} = 1$ mg/min, $K_m = 15$ mg, $\sigma_q^2=0.0002$, $\sigma_c^2 = 0.00003$, and $t= 5$, $10$, $15$, $20$, $25$, $30$, $40$, $50$, $60$, $90$, $120$, $150$, $180$, $230$, $290$, $340$, and $390$ min. There are 17 observation time points, and the corresponding amounts of drug $Q_k$ and concentrations in plasma $c_k$ are generated from equation (\ref{discretestate}) and (\ref{discreteobserved}) respectively. The filtering estimate of $Q_k$ is $Q_{k|k}$, which is calculated from those observed values of concentrations by EKF and DMF respectively. A plot of the observed $C(t)$ and the estimated $Q_{k|k}$ versus time is given in Figure  \ref{estimationqq}.

The mean absolute error (MAE) is defined as
\begin{equation}
\text{MAE} = \frac{1}{n} \sum_{k=1}^{n} \mid Q_k - Q_{k|k}  \mid,
\end{equation}
which is used to measure the accuracy of estimates, where $n$ is the number of observations.

The simulated investigation is repeated 200 times, and the MAE of each simulation is calculated. A quantile analysis is applied to those observed MAEs, and the results are reported in Table \ref{predictions}, where DMF and EKF indicate the results are given by DMF and EKF respectively. And RD is the relative difference between the values of DMF and EKF, i.e.
\begin{equation}
RD= \frac{\text{MAE of EKF}-\text{MAE of DMF}}{\text{MAE of DMF}}.
\end{equation}
It is found that the $0.95$ quantile of MAEs given by DMF is smaller than the $0.05$ quantile given by EKF.
It can be concluded that the errors of estimates given by DMF is much smaller than their counterparts  given by EKF.  This result can be regarded as another evidence  to support the argument in \cite{overgaard2007}, where EKF is found to be inadequate for nonlinear modeling in PK/PD data analysis.

% table of predictions of drug in GI
\begin{table}
\caption{MAEs of estimates for drug in GI tract by DMF and  EKF within 200 simulations.}\label{predictions}
\begin{center}
\begin{tabular*}{\textwidth}{@{\extracolsep{\fill}}lrrrrrrrrr}
\hline
Quantiles	&	0.0500 	&					0.3000 	&			0.5000 	&	0.6000 	 &	0.7000 	&	 0.8000 	&	0.9000 	&	0.9500 	\\
\hline
DMF	&	0.0233 	&					0.0335 	&			0.0399 	&	0.0418 	&	 0.0448 	&	0.0487 	 &	0.0546 	&	0.0591 	\\
EKF	&	0.0600 	&					0.0738 	&			0.0797 	&	0.0825 	&	 0.0863 	&	0.0908 	 &	0.0990 	&	0.1021 	\\
RD	&	1.5747 	&					1.2042 	&			0.9997 	&	0.9735 	&	 0.9246 	&	0.8647 	 &	0.8117 	&	0.7288 	\\
\hline
\end{tabular*}
\end{center}
\end{table}

\section{Estimation of parameters}

\subsection{Criterion of estimation}

MLE based on the extended Kalman filter (EKF) is used to estimate  the parameters in SDE models by several authors, such as  \cite{kristensen2005,tornoe2005,overgaard2007,overgaard2005,moller2010,picchini2010} and the references therein. On the other hand,
MLE is often inefficient and biased for finite sample.
As the sample size is limited  in this paper,  an alternative criterion of estimation is adopted to estimate the unknown parameters.

For a particular parameter $\theta$, the filtering estimate of $Q_k$ is denoted as $Q_{k|k}(\theta)$, which is a function of $\theta$. Substitute $Q_{k-1}$ with $Q_{k-1|k-1}(\theta)$ in equation (\ref{discreteobserved}), and simulate $M$ observations of $C_k$ from  this equation, denoted as $c_{1,k},c_{2,k}, \cdots, c_{M,k}$. Let
\begin{equation}
\rho(c_k,\theta)= \sum_{j=1}^{M} \mid c_k - c_{j,k} \mid,
\end{equation}
where $c_k$ is the observed value of $C_k$ at time $t_k$. Let
\begin{equation} \label{objectivefunction}
L(\theta) = \sum_{k=1}^{n} \rho(c_k,\theta),
\end{equation}
which is the loss function to be used in the following sections.
The parameter $\hat{\theta}$ which satisfies
\begin{equation}\label{optimization}
\hat{\theta} = arg \min_{\theta \in \Theta} L(\theta)
\end{equation}
is used as the estimator of  $\theta^*$ which generates those observed data. $\hat{\theta}$ is a simulation-based quasi-robust estimator, which is insensitive to departures from underlying assumptions.

\subsection{Optimization procedure}

The objective function (\ref{objectivefunction})  is a nonlinear function, where $c_{j,k}$ is simulated from equation (\ref{discreteobserved}) based on the filtering estimate of $Q_{k}$, such that $\hat{\theta}$ can not be computed explicitly. In order to solve the nonlinear optimization problems in PK/PD data analysis, a quasi-Newton method based on BFGS updating formula is adopted by several authors, such as \citep{kristensen2005,kristensen2003} and the references therein, where the gradient of the objective function is approximated by a set of finite difference derivatives. This algorithm can be shown to converge to a possible local minimum \citep{kristensen2003}.

In order to avoid the attraction of local minimum, a genetic algorithm is proposed in this paper. The steps of this algorithm are as follows:
\begin{enumerate}
\item Start. Generate random population of $S$ parameters $\theta_i \in \Theta$ , where  $i=1,2,\cdots,S$.
\item Fitness. Evaluate the fitness of each parameter in the population with $L(\theta)$, where the smaller $L(\theta)$, the better fitness.
\item New population. Create a new population by repeating following steps until the new population is complete.
\begin{enumerate}
\item Selection. $[\alpha S_{pop}]$ parameters are selected from the population according to their fitness.  $[x]$ indicates the largest integer which is less than $x$, $S_{pop}$ is the size of the current population, and $\alpha$ is the proportion of parameters selected to be new population.
\item Crossover. Denote the selected parameters $\theta_1,\theta_2,\cdots,\theta_C$, which are sorted in  increasing order according to their values of $L(\theta)$.  The crossover probability for the ith parameter is chosen to be
\begin{equation}
p_i = \frac{L_{i+1}-L_{i}}{L_{C}-L_{1}},
\end{equation}
where $i=1,2,\cdots,C-1$,  and $L_i=L(\theta_i)$. $\theta_1,\theta_2,\cdots,\theta_{C-1}$ are randomly selected to cross over according to those probabilities.  Suppose $\theta_i$ and $\theta_j$ are selected to cross over in the $r$th run, where $L_i \le L_j$, $r=1,2,\cdots,R$, and $\theta_i = \left( \theta_{i,1},\theta_{i,2},\cdots,\theta_{i,6} \right)$. Two new parameters are generated in the following way,
    \begin{eqnarray}
    \theta_{r}^{new} &=& \theta_i + \left( \theta_i-\theta_j \right) \times weight \times temperature,\\
    \widetilde{\theta}_{r}^{new} &=& \theta_i - \left( \theta_i-\theta_j \right) \times weight \times temperature,\\
    weight &=& \frac{L_i}{L_i+L_j},
    \end{eqnarray}
where $temperature$ is a parameter to control the speed of convergence.
If the jth element $\theta_{r,j}^{new}$ of the new parameter $\theta_{r}^{new}=\left(\theta_{r,1}^{new},\theta_{r,2}^{new},\cdots,\theta_{r,6}^{new}\right)$ is lager than the upper bound $\overline{\theta}_j$, or smaller than the lower bound $\underline{\theta}_j$, then set
\begin{equation}
\theta_{r,j}^{new} =  \frac{3 \theta_{i,j} + \overline{\theta}_j}{4}, \quad \text{or} \quad
\theta_{r,j}^{new} =  \frac{3 \theta_{i,j} + \underline{\theta}_j}{4}.
\end{equation}
Which is also true for $ \widetilde{\theta}_{r}^{new} $.

\item Mutation. Another $S$ parameters are generated randomly from the set $\Theta$, denoted as $\theta_{1}^{M}$, $\theta_{2}^{M}$,$\cdots$, $\theta_{S}^{M}$.

\item Accepting. Evaluate the values of loss function at $ \theta_1^{},\theta_2,\cdots $, $\theta_C $,  $\theta_1^{new}$, $\theta_2^{new}$, $\cdots$,$\theta_R^{new}$, $\widetilde{\theta}_1^{new}$, $\widetilde{\theta}_2^{new}$, $\cdots$, $\widetilde{\theta}_R^{new}$, $\theta_{1}^{M}$, $\theta_{2}^{M}$,$\cdots$, $ \theta_{S}^{M}$, and select the front $[\alpha (C+2R+M)]$ parameters as the new population.
\end{enumerate}

\item Replace. Use new generated population for a further run of algorithm.

\item Test. Denote the populations in the $i$th and $(i+1)$th generations as $\left\{ \theta^{i}_{1},\theta^{i}_{2},\cdots,
    \theta^{i}_{n_i} \right\}$ and $\left\{ \theta^{i+1}_{1},\theta^{i+1}_{2},\cdots,
    \theta^{i+1}_{n_{i+1}} \right\}$. Set a series of probabilities $0 < \alpha_1 < \alpha_2< \cdots < \alpha_a < 1$ and find those corresponding quantiles of these two populations, denoted as
      $\left\{ \theta^{i}_{\alpha_1},\theta^{i}_{\alpha_2},\cdots,
    \theta^{i}_{\alpha_a} \right\}$  and $\left\{ \theta^{i+1}_{\alpha_1}\right.$, $\theta^{i+1}_{\alpha_2}$, $\cdots$,
    $\left. \theta^{i+1}_{\alpha_{a}} \right\}$. Let
    \begin{equation}
     EC=\frac{1}{a}
     \sum_{j=1}^{6}\sum_{k=1}^{a}\left| \frac{\theta^{i}_{\alpha_k,j}-\theta^{i+1}_{\alpha_k,j}}
      {\theta^{i}_{\alpha_k,j}} \right|,
    \end{equation}
    the end condition is chosen to be $EC \le 0.00001$, or the number of loops is beyond 100. If the end condition is satisfied, stop and  return the best solution in current population. The estimator $\hat{\theta}$ is the mean of the last population.

\item Loop. Go to step 2.

\end{enumerate}

Let $\theta_1^i$ and $\theta_1^{i+1}$ be the first parameters in the $i$th and $(i+1)$th generations, which must satisfy
\begin{equation}
0 \le L\left(\theta_1^{i+1}\right) \le L\left(\theta_1^{i}\right) \le L\left( \theta_1^{1} \right),
\end{equation}
which ensures the convergence of algorithm. The mutation step reduces the risk of attraction of local minima.
The proposed genetic algorithm needs not to approximate the gradients with a set of finite difference derivatives, and the burden of programming is much less than the quasi-Newton method based on the BFGS updating formula.

\subsection{Estimates comparison}
Set $Q_0 = 5$ mg, $c_0 = 0$ mg/l, $C_L = 0.05$ l/min, $V=5$ l, $V_{max} = 1$ mg/min, $K_m = 15$ mg, $\sigma_q^2=0.0002$, $\sigma_c^2 = 0.00003$, and $t= 5,10$, $15,20,25,30,40$, $50,60,90,120,150$, $180,230,290,340$, and $390$ min. There are 17 time points, and those observations are generated from (\ref{discretestate}) and (\ref{discreteobserved}).  Parameter $\theta=\left( V_{max}, K_{m}\right.$, $V$, $C_{L}$, $\sigma_q^2$, $\left. \sigma_c^2 \right)$ will be estimated from the limited observations by the proposed estimators based on EKF and DMF respectively.

The simulated experiment  is repeated 200 times in this paper, and the vector of parameters is estimated in each simulation, where $temperature =0.75$, the proportion $\alpha$ used to select new populations is $0.05$, and the quantiles used to construct the end condition are $0.2,0.4,0.5,0.6$, and $0.8$. A quantile analysis is applied to those estimated parameters, and the results are reported in Table \ref{estimations}. There are 11 quantiles reported in this table, including $0.05$, $0.1$, $0.2$, $0.3$, $0.4$, $0.5$, $0.6$, $0.7$, $0.8$, $0.9$ and $0.95$. The real value of each parameter is given in the line \textit{Real}. And entries in the first row of each quantile are the estimated parameters given by the algorithm based on DMF, and the second row are those estimates given by the method based on EKF respectively.

% table of parameters estimates
\begin{table}[h]
\caption{Quantile analysis of parameters estimated by DMF and EKF within 200 simulations.}\label{estimations}
\begin{center}
\begin{tabular*}{\textwidth}{@{\extracolsep{\fill}}rrrrrrr}
\hline
        &   $V_{max}$ & $Km$       & $V$        & $C_{L}$   & $\sigma^2_{q}$ & $\sigma^2_c$ \\
\hline
\textit{Real}	&	1.0000 	&	15.0000 	&	5.0000 	&	0.0500 	&	0.00020 	 &	 0.00003 	\\
0.05 	&	0.5480 	&	7.2686 	&	4.0427 	&	0.0423 	&	0.00007 	&	0.00001 	 \\
 	&	0.3939 	&	8.2121 	&	2.6350 	&	0.0389 	&	0.00005 	&	0.00001 	 \\
0.10 	&	0.6438 	&	8.8172 	&	4.2487 	&	0.0446 	&	0.00012 	&	0.00001 	 \\
 	&	0.4556 	&	9.4951 	&	2.8937 	&	0.0442 	&	0.00008 	&	0.00001 	 \\
0.20 	&	0.7142 	&	10.7491 	&	4.4982 	&	0.0470 	&	0.00018 	&	 0.00001 	\\
 	&	0.5307 	&	10.8259 	&	3.2882 	&	0.0486 	&	0.00016 	&	0.00002 	 \\
0.30 	&	0.7792 	&	12.0278 	&	4.7213 	&	0.0494 	&	0.00029 	&	 0.00002 	\\
 	&	0.5951 	&	12.3268 	&	3.5474 	&	0.0507 	&	0.00023 	&	0.00002 	 \\
0.40 	&	0.8425 	&	13.5821 	&	4.8500 	&	0.0505 	&	0.00035 	&	 0.00002 	\\
 	&	0.6460 	&	13.4156 	&	3.7390 	&	0.0528 	&	0.00028 	&	0.00002 	 \\
0.50 	&	0.8974 	&	14.2116 	&	4.9408 	&	0.0525 	&	0.00041 	&	 0.00002 	\\
 	&	0.6896 	&	14.1414 	&	4.0071 	&	0.0554 	&	0.00034 	&	0.00003 	 \\
0.60 	&	0.9643 	&	15.0688 	&	5.0959 	&	0.0540 	&	0.00048 	&	 0.00002 	\\
 	&	0.7320 	&	14.9648 	&	4.1787 	&	0.0579 	&	0.00041 	&	0.00003 	 \\
0.70 	&	1.0299 	&	16.2975 	&	5.2423 	&	0.0562 	&	0.00056 	&	 0.00003 	\\
 	&	0.7789 	&	16.1194 	&	4.4300 	&	0.0602 	&	0.00048 	&	0.00003 	 \\
0.80 	&	1.0819 	&	17.0248 	&	5.4162 	&	0.0586 	&	0.00067 	&	 0.00003 	\\
 	&	0.8459 	&	17.4575 	&	4.7500 	&	0.0625 	&	0.00058 	&	0.00004 	 \\
0.90 	&	1.1604 	&	18.1916 	&	5.7446 	&	0.0619 	&	0.00082 	&	 0.00004 	\\
 	&	0.9895 	&	18.7287 	&	4.9809 	&	0.0662 	&	0.00066 	&	0.00004 	 \\
0.95 	&	1.2349 	&	18.8432 	&	5.9881 	&	0.0655 	&	0.00090 	&	 0.00004 	\\
 	&	1.0518 	&	18.9551 	&	5.3283 	&	0.0718 	&	0.00071 	&	0.00004 	 \\
\hline
\end{tabular*}
\end{center}
\end{table}

The mean absolute error of estimated parameters  is defined as
\begin{equation}
\text{MAEP}=\frac{1}{6N} \sum_{i=1}^{N} \sum_{j=1}^{6} \mid \hat{\theta}_{i,j} - \theta^{*}_{j} \mid,
\end{equation}
where $\hat{\theta}_{i,j}$ is the jth element of the ith estimated parameter in $N$ simulations, and $\theta_j^*$ is the jth element of the real parameter which generates the observed data. The MAEP of EKF is 0.7070 among 200 simulations,  and its counterpart of DMF is 0.5929, where the later is  $16.13\%$ less than the former, which indicates that the estimator based on DMF is better than the one based on EKF with respect to MAEP.

The algorithms proposed in this paper are programmed with Matlab R2009, which run on a personal computer with an Intel(R) Core(TM)2 Duo CPU E7500, whose main frequency is double 2.93 GHz.

\section{Conclusions and discussions}

A density-based Monte Carlo filter  is proposed to estimate the unobservable variables in a nonlinear stochastic differential equation model, and a simulation-based quasi-robust estimator is proposed to estimate the unknown pharmacokinetic parameters in this model. A genetic algorithm is proposed to solve the optimization problem in the estimation procedure. The performance of the proposed filter is compared with the extended Kalman filter, and it is found that DMF is more efficient than EKF in the simulation investigations.

Further research possibilities are mainly in three directions. First of all, other nonlinear filters can be applied in the analysis of PK/PD data.
Several nonlinear filters are used to estimate the unobservable variables in state-space models, including the Gaussian sum filter, the numerical integration filter, the importance sampling filter, the rejection sampling filter, and the density-based Monte Carlo filter. It should be possible to determine the optimal filtering algorithm for a particular PK/PD  model. The second direction concerns the estimation criterion that can be used in the analysis of PK/PD data. It is often stated that MLEs are biased for finite sample, while the sample size in PK/PD data analysis is often limited. The third direction is the algorithm to be used in the procedure of optimization. The efficiency of optimizing algorithms should be taken into account in PK/PD data analysis.

\section*{Acknowledgements}
This project is supported by NSF of China under Grant 30 973 586.
And Guanghui Huang is also supported by the Fundamental Research Funds for the Central Universities of China under Grant CDJZR10 100 007.

\appendix

\section{Algorithm of EKF}\label{appekf}
Suppose $y_k$ is the value of observable variable at time $t_k$, and $\alpha_k$ is the unobservable state variable at time $t_k$, which satisfy
\begin{eqnarray}
y_k & = & h_k (\alpha_k, \epsilon_k),\\
\alpha_k & = & g_k (\alpha_{k-1},\eta_k),
\end{eqnarray}
where $\epsilon_k$ and $\eta_k$ are two independent disturbances at time $t_k$.  $h_k(\cdot)$ and $g_k(\cdot)$ are two nonlinear functions, which can be approximated with  first order Taylor series expansions
\begin{eqnarray}
y_k & \approx &  h_{k|k-1} + Z_{k|k-1} (\alpha_k-\alpha_{k|k-1})
   + S_{k|k-1}\epsilon_k, \\
\alpha_k & \approx &
  g_{k|k-1}+ T_{k|k-1} (\alpha_{k-1} - \alpha_{k-1|k-1}) + R_{k|k-1} \eta_{k},
\end{eqnarray}
where
\begin{eqnarray*}
h_{k|k-1} & = & h_k(\alpha_{k|k-1},0),\\
Z_{k|k-1} & = &
 \left.\frac{ \partial h_k(\alpha_k,\epsilon_k)}{\partial \alpha_k}  \right| _{(\alpha_k,\epsilon_k)=(\alpha_{k|k-1},0),}\\
 S_{k|k-1} & = &
  \left. \frac{\partial h_{k}(\alpha_k,\epsilon_k)}{\partial \epsilon_k} \right|_{(\alpha_k,\epsilon_k)=(\alpha_{k|k-1},0),}\\
g_{k|k-1} & = &
  g_{k}(\alpha_{k-1|k-1},0),\\
T_{k|k-1} & = &
 \left.\frac{ \partial g_k(\alpha_{k-1},\eta_k)}{\partial \alpha_{k-1}} \right| _{(\alpha_{k-1},\eta_k)=(\alpha_{k-1|k-1},0),}\\
 R_{k|k-1} & = &
  \left. \frac{\partial g_{k}(\alpha_{k-1},\eta_k)}{\partial \eta_k} \right|_{(\alpha_{k-1},\eta_k)=(\alpha_{k-1|k-1},0).}
\end{eqnarray*}
EKF is given by the following algorithm:
\begin{eqnarray}
\alpha_{k|k-1} & = &
      g_{k|k-1},\\
\Sigma_{k|k-1} & = &
      T_{k|k-1} \Sigma_{k-1|k-1} T'_{k|k-1} + R_{k|k-1} Q_{k} R'_{k|k-1},\\
y_{k|k-1} & = &
      h_{k|k-1},\\
F_{k|k-1} & = &
      Z_{k|k-1} \Sigma_{k|k-1} Z'_{k|k-1} + S_{k|k-1} H_t S'_{k|k-1},\\
M_{k|k-1} & = &
      Z_{k|k-1} \Sigma_{k|k-1},\\
K_k & = &
     M'_{k|k-1} F^{-1}_{k|k-1},\\
\Sigma_{k|k} & = &
     \Sigma_{k|k-1} - K_k F_{k|k-1} K'_{k},\\
\alpha_{k|k} & = &
     \alpha_{k|k-1} + K_{k} \left( y_k - y_{k|k-1} \right),
\end{eqnarray}
where $\Sigma_{0|0}=0$, $\alpha_{0|0}= Q_0$, and $Q_k = H_{k}= t_k - t_{k-1}$ for $k=1,2,\cdots,n$ in this paper. The details of EKF can be found in \cite{tanizaki1996}.

\section{Algorithm of DMF}\label{appdmf}
Denote the collection of state-vector as
\begin{equation}
A_t = \left\{ q_0, q_1, \cdots, q_t  \right\},
\end{equation}
where $q_k$ is the value of unobservable variable at time $t_k$, $k=1,2,\cdots,n$. The joint density function of $(Y_t,A_t)$ is
\begin{equation}
p\left( Y_t, A_t \right)=p\left(A_t\right)p\left(Y_t|A_t\right),
\end{equation}
where $p\left(A_t\right)$ and $p\left(Y_t|A_t\right)$ are
\begin{eqnarray}
p\left(A_t\right) & = &
  p\left(q_0\right)\prod_{s=1}^{t} p\left(q_s|q_{s-1}\right),\\
p\left(Y_t|A_t\right) & = &
 \prod_{s=1}^{t}p\left(c_s|q_{s-1}\right),
\end{eqnarray}
where $p\left(q_s|q_{s-1}\right)$ and $p\left(c_s|q_{s-1}\right)$ are obtained from (\ref{discretestate}) and (\ref{discreteobserved}) respectively. The filtering density function is given by
\begin{equation}
p\left(q_s|Y_{s}\right) = \frac{\int p\left(q_s,A_{s-1},Y_s\right) \mathrm{d}A_{s-1} }
                          {\int p\left(A_{s}\right)
                           p\left(Y_s|A_{s}\right) \mathrm{d}A_s},
\end{equation}
such that the filtering estimate of the state variable is given by
\begin{equation}
Q_{t|t}   =  \mathrm{E}\left[Q_t | Y_t \right] = \frac{\int q_t p\left(Y_t|A_t\right) p\left(A_t\right) \mathrm{d}A_t}{\int p\left(Y_t|A_t\right) p\left(A_t\right) \mathrm{d}A_t}.
\end{equation}
Generating random draws of $A_t$ from $p\left(A_t\right)$, the filtering estimate based on the Monte Carlo technique is give by
\begin{eqnarray}
Q_{t|t} & = & \frac
              {\frac{1}{N}\sum_{i=1}^{N} Q_{i,t} p\left(Y_t|A_{i,t}\right)}
               {\frac{1}{N}\sum_{i=1}^{N}  p\left(Y_t|A_{i,t}\right)}\nonumber\\
              & = &
               \frac
              {\sum_{i=1}^{N} Q_{i,t} \prod_{s=1}^{t} p\left(y_s|q_{i,s-1}\right)}
               {\sum_{i=1}^{N}  \prod_{s=1}^{t} p\left(y_s|A_{i,s-1}\right)},
\end{eqnarray}
where $A_{i,t}$ is the collection of random draws for the ith generated path, i.e.
\begin{equation}
A_{i,t} = \left\{ Q_{i,0}, Q_{i,1}, \cdots, Q_{i,t}\right\}.
\end{equation}

Denote
\begin{eqnarray*}
\omega_{j,k}=\frac{\prod_{s=1}^{k} p\left(y_s | Q_{j,s-1}\right)}
             {\sum_{j=1}^{N} \prod_{s=1}^{k} p\left( y_s | Q_{j,s-1}\right)},
\end{eqnarray*}
then we have
\begin{equation*}
\omega_{j,k} = \frac{p\left(y_k | Q_{j,k-1}\right) \omega_{j,k-1}}{\sum_{j=1}^{N}p\left(y_k | Q_{j,k-1}\right) \omega_{j,k-1}},
\end{equation*}
where
\begin{equation*}
\sum_{j=1}^{N}\omega_{j,k}=1,
\end{equation*}
and $\omega_{j,0}=1/N$. The DMF of $Q_k$ is given by
\begin{equation}
Q_{k|k} = \sum_{j=1}^{N} Q_{j,k}\omega_{j,k}.
\end{equation}
The details of DMF can be found in \cite{tanizaki1996}.

\bibliographystyle{elsarticle-harv}



\end{document}